\documentclass[%
singlecolumn
]{revtex4-2}
\usepackage{graphicx}
\usepackage{subcaption}
\usepackage{pstricks}
\usepackage{color}
\usepackage{amsmath}
\usepackage{soul}
\usepackage{bm}

\usepackage{bbm}

\begin{document}

\def\vsm{v_{\tt SM}}
\def\uu{s}
\def\sw{s_{\tt w}}
\def\cw{c_{\tt w}}
\def\gz{g_{\tt z}}
\def\gd{g_{\tt D}}

\preprint{APS/123-QED}

\title{Neutrino masses and self-interacting dark matter with mass mixing $Z-Z^\prime$ gauge portal}

\author{Leon M.G. de la Vega }%
 \email{leonm@estudiantes.fisica.unam.mx}
\affiliation{%
 Instituto de F\'{\i}sica, Universidad Nacional Aut\'onoma de M\'exico, A.P. 20-364, Ciudad de M\'exico 01000, M\'exico.}%
\author{Eduardo Peinado}%
 \email{epeinado@fisica.unam.mx}
\affiliation{%
 Instituto de F\'{\i}sica, Universidad Nacional Aut\'onoma de M\'exico, A.P. 20-364, Ciudad de M\'exico 01000, M\'exico.}%
\author{Jos\'e Wudka }%
 \email{jose.wudka@ucr.edu}
\affiliation{%
 Department of Physics and Astronomy, UC Riverside, Riverside, California 92521-0413, USA.}%
\date{\today}

\begin{abstract}
New light gauge bosons can affect several low-energy experiments, such as atomic parity violation or colliders. Here, we explore the possibility that a dark sector is charged under a new $U(1)$ gauge symmetry, and the portal to the Standard Model is through a $Z-Z'$ mass mixing. In our approach, breaking the new gauge symmetry is crucial to generate neutrino masses. We investigate the parameter space to reproduce neutrino masses, the correct dark matter relic abundance, and to produce the observed core-like DM distribution in galactic centers.
\end{abstract}

\maketitle
                                                                                                                                                                                                                                   

\section{Introduction}
The existence of dark matter and the non-zero neutrino masses are not explained within the Standard Model (SM) of particle physics. Dark matter (DM) is a component of the universe that accounts for $\sim 27 \%$ of its total matter-energy density \cite{Planck:2018vyg}. No particle, fundamental or composite, in the SM can account for it. A possible way to incorporate dark matter into the SM framework is to extend the gauge group $G_{SM}=SU(3)_C\times SU(2)_L\times U(1)_Y$ to include a dark gauge sector, under which the DM candidate is charged. One of the simplest ways to extend the SM is with an extra abelian gauge symmetry $U(1)_D$, which enlarges the gauge boson content of the SM. The dark sector of such a theory may communicate with the SM particles via the kinetic mixing term of the abelian subgroups $U(1)_Y$ and $U(1)_D$, the mass mixing among the neutral gauge bosons, or the scalar sector. The connection through the kinetic mixing is a popular and widely explored paradigm known as the {\it dark photon} \cite{Holdom:1985ag,Fayet:1980ad,Fayet:1980rr,Fayet:1990wx,Okun:1982xi,Georgi:1983sy}, where the dark gauge boson acquires vector couplings to the SM fermions. In the mass mixing case~\cite{Davoudiasl:2012ag,Davoudiasl:2014kua}, the dark gauge boson acquires both vector and axial vector couplings to the SM fermions, leading to possible signatures in parity-violation experiments. This new gauge boson can provide a viable communication channel between the DM and the SM, leading to the correct dark matter relic density through the freeze-out while avoiding direct-detection constraints \cite{Jung:2020ukk}.\\On the other hand, many neutrino mass generation mechanisms explain the lightness of Majorana neutrinos, compared to the rest of SM fermions. Among them, a popular class of models is the so-called {\it seesaw} mechanism, where a large mass scale suppresses the electroweak scale in the neutrino masses, giving rise to small neutrino masses. A popular type of these models requires the introduction of right handed (RH) neutrinos, singlets of the SM. Different mass models may be obtained, depending on the mass terms present in the lagrangian after the breaking of $G_{SM}$ and any other additional symmetries in the model, namely, the type-I seesaw \cite{Minkowski:1977sc,Gell-Mann:1979vob,Mohapatra:1979ia,Yanagida:1979as,Schechter:1980gr}, the linear seesaw \cite{Wyler:1982dd,Akhmedov:1995ip,Akhmedov:1995vm} or the inverse seesaw\cite{Mohapatra:1986aw,Mohapatra:1986bd}. Each of these models results in different possible values for the heavy neutrino masses and active-sterile neutrino mixings.  In this work, we study the SM extended with an anomaly-free $U(1)_D$ gauge symmetry. The fermions, charged under the new gauge symmetry, will be identified with the right-handed neutrinos and the dark matter candidate. An extra Higgs doublet, charged under the dark gauge symmetry, generates a mass mixing among the dark gauge boson and the electroweak neutral gauge boson. The right-handed neutrinos' and Higgs fields' charges shape the neutrino seesaw matrix \cite{Heeck:2012bz}. Dark matter is connected to the SM matter fields through the neutral gauge boson mass mixing, opening up viable thermal freeze-out channels and signatures in direct detection experiments.

\section{The model}
\begin{table}[!h]
\label{tab:linear}$
\begin{array}{||l|l|l|l|l|l|l|l|l|l||}
\hline         & L    & N & N^\prime& F & H_1 & H_2 & \phi & \chi_L  & \chi_R^c\\\hline
SU(2)_L & 2    & 1 & 1    &1                                    & 2    & 2    & 1    &1& 1 \\\hline
U(1)_Y  & -1/2 & 0 & 0      &0                                  & 1/2  & 1/2  & 0 &0 & 0 \\\hline
U(1)_D  & 0    & 1 & -1       &0                                & 0    & 1    & -1 &Q_{\tt D} &-Q_{\tt D}\\ \hline
\end{array}$
\caption{Matter content of the dark matter with mass mixed $U(1)_D$ gauge boson. The model contains two dark charges, we have chosen to absorb one of them into the definition of the dark gauge coupling, leaving the dark matter charge $Q_D$ free.}
\label{tab:mattercontent}
\end{table}
We consider the model for the mass mixing of a new gauge boson with the $Z$ boson, described by Table~\ref{tab:mattercontent} \footnote{Note that the charge assignment under the new $U(1)_D$ for $N$ and $N^\prime$ could take values different value as far as they remain vector-like, but it is fixed by convenience. To have this, we add to the Standard Model(SM) a new "\textit{dark}" gauge symmetry $U(1)_D$, a scalar $SU(2)_L$ doublet $H_2$ charged under the new symmetry, and a scalar singlet $\phi$ to trigger the symmetry breaking. The right handed neutrinos are charged under $U(1)_D$. Therefore, at least two different sets of vector-like fermions $N$ and $N^\prime$ are needed to cancel the anomalies. We also need the inclusion of an extra fermion singlet under the new symmetry, $F$, to break the lepton number.}. The dark sector (stable after the SSB) consists of a vector-like pair of fermions $\chi_L$ and $\chi_R$. Fermions charged under $U(1)_D$ transform trivially under the SM gauge symmetry, guaranteeing the cancellation of all mixed SM-dark anomalies. For each charged fermion under $U(1)_D$, there is a  fermion with an opposite charge, such that the pure $U(1)_D$ and the $U(1)_D$-gravity anomalies vanish. The right-handed neutrinos $N,~N^\prime,~F$, participate in the seesaw mechanism, with their $U(1)_D$ charges shaping the seesaw mass matrix. The scalar sector will induce mass mixing among the electroweak and dark neutral bosons, linking the SM fermions with the dark sector.  The $\chi$ fields can act as a dark matter candidate, interacting with SM fields through the mass-mixed dark gauge boson.
In this way, we show that the $U(1)_D$ can drive the phenomenology of neutrino and dark matter. The neutral gauge boson mixing will impact the quark and lepton physics, such as parity violation in polarized electron-nucleon and electron-electron scattering.
\subsection{Neutrino sector}
The RH neutrinos are charged under the $U(1)_D$.  To generate the Yukawa Lagrangian, their charges must match that of the $H_2$ Higgs doublet. To avoid extra Goldstone bosons, in the scalar sector we must have a term such as $H_2 H_1^\dagger \phi$ or $H_2 H_1^\dagger \phi^2$. From these two conditions, we conclude that the charge of $\phi$ equals one of the RH neutrinos charges. The two RH neutrinos N and N' have a Dirac mass term. In contrast, there is no way to generate a Majorana mass for any of those fields through the $\phi$ field. The only way to do so is to include an extra fermion with no $U(1)_D$ charge. In this way, the Lagrangian density of the neutrino sector is
\begin{equation}
    \mathcal{L}_{\nu}=Y^\nu_1 \overline{L}\tilde{H}_1 F + Y^\nu_2 \overline{L}\tilde{H}_2 N + M_1 \overline{N^c}N' + Y^N \overline{N^c} F \phi +Y^{N^\prime} \overline{N^{\prime C}} F \phi^* + M_F \overline{F^C}F +h.c.
\end{equation}
After spontaneous symmetry breaking (SSB)  the resulting neutrino mass matrix in the $(\nu_L, N,N',F)$ basis is
\begin{equation}
    M=
\begin{pmatrix}
0               & m^D_2   & 0                       &m^D_1 \\
(m^D_2)^T  & 0              & M_1                     &Y^N v_\phi\\
0               &(M_1)^T         & 0                       &Y^{N^\prime} v_\phi \\
(m^D_1)^T  & (Y^N)^T v_\phi & (Y^{N^\prime})^T v_\phi & M_F
\end{pmatrix},
\end{equation}
where $m^D_i=Y^\nu_i v_i$ are the Dirac mass matrices. The light neutrino mass matrix is given by
\begin{equation}
    m_{light}=(m^D_1)^T
    \alpha m^D_1 +(m^D_2)^T \beta m^D_1 + (m^D_1)^T \delta m^D_2 + (m^D_2)^T \epsilon m^D_2,
\end{equation}
where the $\alpha,\beta,\delta,\epsilon$ matrices are defined as
\begin{equation}
    \begin{split}
        \alpha =& (M_F + Y^N(M_1^T)^{-1}(Y^{N^\prime})^T v_{\phi}^2+Y^{N^\prime} (M_1)^{-1}(Y^{N})^T v_{\phi}^2)^{-1}, \\ \\
        \beta=& ((Y^N)^Tv_{\phi})^{-1} + ((Y^N)^Tv_{\phi})^{-1}M_1(M_1^T)^{-1}[Id+M_1^T(Y^Nv_{\phi})^{-1}( Y^{N\prime}v_{\phi}M_1^{-1} (Y^Nv_{\phi})^{-1}-M_F )((Y^{N\prime}v_{\phi})^T)^{-1} ]^{-1}, \\ \\
        \delta=&-( M_1^T  )^{-1} (Y^{N\prime}v_{\phi})^T[M_F + Y^N(M_1^T)^{-1}(Y^{N^\prime})^T v_{\phi}^2+Y^{N^\prime} (M_1)^{-1}(Y^{N})^T v_{\phi}^2]^{-1}, \\ \\
        \epsilon=&-(Y^Nv_\phi)^{-1}[Y^{N\prime}v_\phi(M_1^T)^{-1}[Id+M_1^T(Y^Nv_{\phi})^{-1}( Y^{N\prime}v_{\phi}M_1^{-1} (Y^Nv_{\phi})^{-1}-M_F )((Y^{N\prime}v_{\phi})^T)^{-1} ]^{-1} +
         M_F \beta],
    \end{split}
\end{equation}
where $Id$ is the Identity Matrix. The minimal field content leading to two massive light neutrinos is $(Nr(N)=2,Nr(N')=2,Nr(F)=2)$.
There are several familiar limits to this framework:
\begin{enumerate}
    \item The type-I seesaw limit can be obtained when ${Y^N,Y^{N'}}\rightarrow 0$ or ${Y^N,m^D_2}\rightarrow 0$ or ${Y^{N'},m^D_2}\rightarrow 0$. The magnitude of light neutrino masses is
    \begin{equation}
        m_\nu\sim \frac{v_1^2}{M_F}.
    \end{equation}
    \item When ${M_F,m^D_1}\rightarrow 0$, the light neutrino masses take the form of the inverse seesaw 
    \begin{equation}
        m_\nu\sim \frac{(m_2^D)^2 Y_{N'}}{M_1 Y_N}.
    \end{equation}
    \item When ${M_F,m^D_2}\rightarrow 0$, the light neutrino masses take a inverse seesaw form
    \begin{equation}\label{seesaw1}
        m_\nu\sim \frac{(m_D^1)^2 M_1}{v_\phi^2 Y_N Y_{N'}}.
    \end{equation}.
     \item When ${Y_N,m^D_1}\rightarrow 0$, the light neutrino masses take the form
    \begin{equation}\label{seesaw2}
        m_\nu\sim \frac{(m_2^D)^2 (Y_{N'})^2 v_\phi^2}{M_1^2 M_F}.
    \end{equation}
    In this case,  there is an extra suppression compared with the inverse seesaw from the light (heavy) scale $v_\phi$ ($M_F$).
\end{enumerate}
We will examine the viability of each limit, depending on the scale of $v_\phi$ indicated by DM phenomenology in section (\ref{sec:neutrinomasspheno}).

\subsection{Dark Sector}
We choose as dark matter candidate a Dirac fermion $\chi=\chi_L+\chi_R$ with mass term
\begin{equation}
    {\cal L}_\chi^{mass}=\frac{1}{2}M_\chi \overline{\chi} \chi .
\end{equation} 
We choose $Q_{\tt D}$ so that Majorana mass terms are forbidden at any order in perturbation theory, with the scalar content of Table~\ref{tab:mattercontent}. The condition to keep the Dirac character of $ \chi $ is
\begin{equation}
    Q_{\tt D} \not= \frac m2,\quad m\in {\mathbbm Z}.
        \label{eq:darkcharge}
\end{equation}

Since $Q_{\tt D}\not=0$, $\chi$ couples to the dark gauge boson $X$; once the gauge symmetry is broken this induces a coupling to both the physical $Z$ boson and the dark photon $Z'$, see Eq. (\ref{eq:gaugecouplings}). For definiteness we will choose $ Q_{\tt D} = 1/3$ that satisfies Eq. (\ref{eq:darkcharge}). A similar dark matter model is described in \cite{Jung:2020ukk}. 

\subsection{Gauge sector}
The $U(1)_D$ charges of the new fields, except for $\chi$, are equal in magnitude. Therefore, we may redefine the gauge coupling and field charges such that $Q=\pm 1$ for these fields. With this in mind, the kinetic terms of the scalar fields in the model defined in Table~\ref{tab:mattercontent} are
\begin{equation}
    \mathcal{L}_{SK}= \sum_{i=1}^2 \left[ (D_\mu H_i)^\dagger (D^\mu H_i) \right] +  \left[ (D_\mu \phi)^\dagger (D^\mu \phi) \right],
\end{equation}
where the covariant derivatives for the $SU(2)_L$ Higgs doublets are
\begin{equation}
D_\mu H_i = (\partial_\mu + \frac{i g}{2}\Vec{\tau}\cdot \Vec{W}_\mu +\frac{ig'}{2}B_\mu + i \gd  Q_i X_\mu )H_i \quad, 
    \end{equation}
with $Q_1=0$ and $Q_2=1$. The corresponding covariant derivative for the $SU(2)_L$ scalar singlet is 
 \begin{equation}   
     D_\mu \phi = (\partial_\mu -  i \gd   X_\mu )\phi \quad .
\end{equation}
After electroweak and dark symmetry breaking, $H_{1,2}$ and $ \phi $ acquire vacuum expectation values and we write
\begin{equation}
    H_1 = \begin{pmatrix}
    H_1^+ \cr
    (v_1 + h_1 + i a_1)/\sqrt{2}
    \end{pmatrix} \,; \qquad
    H_2 = \begin{pmatrix}
    H_2^+ \cr
    (v_2 + h_2 + i a_2)/\sqrt{2}
    \end{pmatrix} \,; \qquad
    \phi = v_\phi + h_\phi + i a_\phi\,.
\label{eq:scalar.fields}
\end{equation}
The SM vacuum expectation value is  $\vsm^2 = v_1^2 + v_2^2$, and we define $ \tan\beta = v_2/v_1 $.
There are five vector bosons, $W^\pm = (W_1\mp i W_2)/\sqrt{2}$ correspond to the usual charged pair with mass $ g \vsm/2 $, one neutral gauge boson, $ A = \sw W_3 + \cw B$ (where $ \sw = \sin(\theta_{\tt w})$ and $ \tan(\theta_{\tt w}) = g'/g$) remains massless. For the remaining fields, let $ \tilde Z = \cw W_3 - \sw B $, then the  mass matrix for $ \{\tilde Z,\,X\}$ becomes
\begin{equation}
    m_{\tilde Z\,X}^2 = \frac14 
    \begin{pmatrix}
    \gz^2 \vsm^2& - 2 v_2^2\, \gd  \gz \cr 
    - 2 v_2^2 \, \gd  \gz  &  4 \gd ^2 (v_2^2 + v_\phi^2 )) 
    \end{pmatrix},
\end{equation}
where $ \gz = \sqrt{g^2 + g'^2}$.
The $\tilde Z$-$X$ mixing angle, $\theta_{\tt X}$, is given by
\begin{equation}
    \tan 2\theta_{\tt X} = \frac{\gz \gd   v_2^2}{\frac{1}{4} \gz ^2\vsm^2- \gd ^2(v_2^2+v_{\phi }^2 )}.
\end{equation}
Denoting the mass eigenstates by $Z$ and $Z'$, the corresponding masses are given by 
\begin{equation}
    M^2_{Z/Z'}=\frac{1}{8} \gz ^2\vsm^2+\frac{1}{2}\gd ^2 (v_2^2+v_{\phi }^2 ) \pm  \\\frac{1}{8} \left\{ \left[\gz ^2\vsm^2-4\gd ^2 (v_2^2+v_{\phi }^2) \right]^2 + \left( 4 \gz  \gd    v_2^2\right)^2   \right\}^{1/2}.
\end{equation}
The $\tilde Z$-$X$ mixing induces a coupling of the $Z' $ to the electroweak neutral current $ J_{\tt NC}$, and a coupling of the $Z$ to the dark current $J_{\tt DC}$ proportional to $ \sin\theta_{\tt X}$; explicitly,
\begin{equation}
    \mathcal{L}_{\tt NC}= - e J^\mu_{\tt EM} A^\mu - Z_\mu (\cos\theta_{\tt X} \frac{\gz }{2} J^\mu_{\tt NC} + \sin\theta_{\tt X} \gd  J^\mu_{\tt DC}) - Z'_\mu (-\sin\theta_{\tt X} \frac{\gz }{2} J^\mu_{\tt NC} + \cos\theta_{\tt X} \gd  J^\mu_{\tt DC}),  
    \label{eq:gaugecouplings}
\end{equation}
where $e=g\sw$. The currents in Eq. (~\ref{eq:gaugecouplings}) are
\begin{equation}
    J_{\tt EM}^\mu=\sum_r Q^{\tt EM}_r \overline{f_r} \gamma^\mu f_r \quad ,\quad J_{\tt NC}^\mu =\sum_r t^3_L(r) \overline{f}_r\gamma^\mu(1-\gamma^5)f_r -2 s_W^2 J^\mu_{\tt EM} \quad ,\quad J^\mu_{\tt DC}=\frac{\gd}{3}\overline{\chi}\gamma^\mu\chi+\gd (\overline{N}\gamma^\mu N -  \overline{N'}\gamma^\mu N'),
    \label{eq:couplings}
\end{equation}
where $Q^{\tt EM}_r$ is the EM charge of the $f_r$ fermion and $t_L^3(r)$ its weak isospin.

\subsection{Scalar Sector.}

The scalar  potential for the model in Table \ref{tab:mattercontent} is given by
\begin{equation}
\label{eq:inversescalarpotential}
\begin{split}
    V_{i} =& \mu_1^2 H_1^\dagger H_1 +\mu_2^2 H_2^\dagger H_2 +\mu_\phi^2 \phi^*\phi + \kappa \phi^* H_1^\dagger H_2 + \lambda_1 (H_1^\dagger H_1)^2 +\lambda_2 (H_2^\dagger H_2)^2 +\lambda_3 (\phi^*\phi)^2  \\
     &+ \lambda_4 (H_1^\dagger H_1)(H_2^\dagger H_2) + \lambda_5 (H_2^\dagger H_2)(\phi^*\phi)+ \lambda_6 (H_1^\dagger H_1)(\phi^*\phi) + \lambda_7 (H_1^\dagger H_2)(H_2^\dagger H_1),
\end{split}
\end{equation}
where the only complex coupling is $\kappa$, however by a field redefinition it can be made real.
From the 10 real scalar degrees of freedom, four goldstone bosons are absorbed in the vector boson masses; the remaining six correspond to a charged pair $ H^\pm $, a pseudoscalar $A$, and three neutral scalars. Using the notation of Eq. (\ref{eq:scalar.fields}) the first three and their masses are given by 
\begin{align}
   & H^+ \sim \sin\beta\, H_1^+ - \cos\beta\, H_2^+ \,,\hspace{3.2cm}  M_{H^+}^2=\frac12 \uu  v_\phi^2  - \frac12 \lambda_7 \vsm^2,  \cr
   & A  \sim v_\phi\sin\beta\, a_1 - v_\phi \cos\beta\, a_2 + \frac12 \vsm \sin(2\beta)\, a_\phi\,, \quad M_A^2= \frac12  \uu v_\phi^2 + \frac{\sin^2(2\beta)}8 \uu   \vsm^2,
\end{align}
while in the $\{h_1,\,h_2,\,h_\phi\}$ basis the CP-even mass matrix is given by
\begin{equation}
    M_{E}^2=\begin{pmatrix}
    2 \lambda_1 v_1^2 -\frac{\kappa v_\phi}{\sqrt{2}} \tan\beta & v_1 v_2 (\lambda_4+\lambda_7)  + \frac{\kappa  v_\phi}{\sqrt{2}}& v_1 v_\phi\lambda_{6}+\frac{\kappa v_2 }{\sqrt{2}}  \\
   v_1 v_2 (\lambda_4+\lambda_7)  + \frac{\kappa  v_\phi}{\sqrt{2}} & 2 \lambda_2 v_2^2 -\frac{\kappa v_\phi}{\sqrt{2}}  \cot\beta         &v_2v_\phi\lambda_5  +\frac{\kappa v_1 }{\sqrt{2}} \\
   v_1 v_\phi\lambda_{6}+\frac{\kappa v_2 }{\sqrt{2}}         &v_2v_\phi\lambda_5  +\frac{\kappa v_1 }{\sqrt{2}}       & 2 \lambda_3 v_\phi^2  -\frac{\kappa v_1 v_2}{\sqrt{2}v_\phi}
    \end{pmatrix}.
\end{equation}
To simplify the expressions we defined
\begin{equation}
    \uu = -\frac{\sqrt{8}\, \kappa}{v_\phi \sin(2\beta)},
\end{equation}
which is positive since in our conventions $ \kappa < 0 $. The scalar potential must be bounded from below, leading to restrictions on the scalar couplings. We have collected these restrictions in Appendix \ref{sec:AppendixScalarStability}. We note the existence of a decoupling limit, where the scalar masses become much heavier than the electroweak scale, save for the Higgs seen at LHC. This limit is achieved when $v_2\rightarrow 0$ and $\kappa\rightarrow 0$, with $\mu_2^2$ setting the heavy scalar scale. The decoupling limit drives the gauge boson mixing to zero, making the $Z'$ and dark matter invisible. 
\section{Phenomenology}
\subsection{General constraints on light dark $Z^\prime$ bosons}
\label{sec:Zprimeconstraints}
The induced couplings of the dark $Z^\prime$ boson to the SM fermions can be probed by a variety of experiments \cite{Davoudiasl:2012ag,Davoudiasl:2014kua,Lindner:2020kko,Dev:2021otb,Ilten:2018crw,Baruch:2022esd}. The observables measured  by those experiments, constrain the parameter space in the $\theta_{\tt X}-M_{Z^\prime}$ plane. The most relevant experimental constraints are the following:
\begin{itemize}
     \item \underline{Atomic Parity Violation}. As noted above, the mass mixing among the SM and $X$ bosons induces a $Z'$ couplings to SM fermions ({\it cf.} Eq. (\ref{eq:couplings})), which inherit the parity violating nature of the SM $Z$ couplings. The $Z'$ parity violating couplings may induce observable effects on low energy experiments, when the mass of the $Z'$ is comparable to the energy scale of the experiment \cite{Bouchiat:2004sp}. This parity-violating interaction of quarks and leptons mediated by the  $Z'$ has been probed in atomic transitions of Yb, Cs, Tl, Pb, and Bi. The measurement of the nuclear weak charge in these experiments can be used to constrain the $Z'$ couplings to the SM fermions as a function of its mass \cite{Davoudiasl:2012ag}. The resulting constraint is approximately \cite{Ilten:2018crw,Baruch:2022esd,delavega2022}
     \begin{equation}
         \sin\theta_{\tt X} \lesssim 5 \times 10^{-5}, ~\text{for}~ M_{Z^\prime}<40 \text{ MeV}\,; \quad \text{and} \quad \frac{\sin\theta_{\tt X}}{ M_{Z^\prime}}\lesssim\frac{10^{-6}}{\text{ MeV}}, ~\text{for}~ 40 \text{ MeV} < M_{Z^\prime} < 100 \text{ GeV}.
     \end{equation}
    \item \underline{Collider searches}. As the $Z'$ couples to the SM fermions, it can be produced in a multitude of collider experiments. The $Z'$-mediated Drell-Yan production of muons in hadron colliders yield some of the strongest constraints on the $Z'$  couplings for masses below $M_Z$. Neutral gauge boson production and decay to leptons, in association with photon production has been searched for in $e^+ e^-$ collisions, at BaBaR \cite{BaBar:2014zli}, LHCb \cite{LHCb:2017trq}, among others. The constraint from colliders in the mass region, $100 \text{ MeV} <M_{Z^\prime}<80 \text{GeV}$, is roughly \cite{Ilten:2018crw,Baruch:2022esd}
    \begin{equation}
         \sin\theta_{\tt X} \lesssim5 \times 10^{-3}.
    \end{equation}
    \item \underline{Beam dump experiments}. The production of neutral bosons in electron bremsstrahlung processes in beam dumps has been probed, for example at the NA64 \cite{NA64:2018lsq}, E141 \cite{Riordan:1987aw}, E137 \cite{Bjorken:1988as} and E774 \cite{Bross:1989mp} experiments. The beam dump limits on $\sin\theta_{\tt X} $ in the $Z'$ gauge boson mass region, $1 \text{ MeV} <M_{Z^\prime}<500 \text{ MeV}$ are roughly \cite{Ilten:2018crw,Baruch:2022esd}
    \begin{equation}
        \sin\theta_{\tt X} \lesssim 3\times 10^{-8} \text{ or } \frac{\sin\theta_{\tt X} }{(M_{Z'}/\text{GeV})^{-1.2}}\gtrsim 3.3\times 10^{-7}   .
    \end{equation}
\end{itemize}

\subsection{Dark matter relic density}
\begin{figure}
\begin{subfigure}{.25\textwidth}
  \centering
  \includegraphics[width=0.7\linewidth]{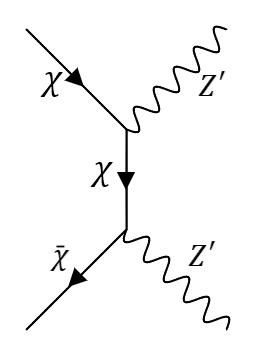}
  \caption{t-channel annihilation into $Z'Z'$}
\end{subfigure}%
\begin{subfigure}{.25\textwidth}
  \centering\vspace{.7cm}
  \includegraphics[width=1.0\linewidth]{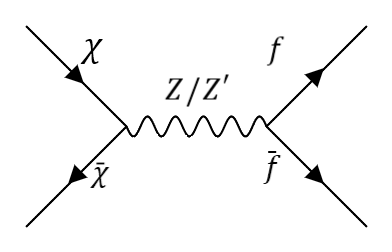}\vspace{.8cm}
  \caption{Z/Z' mediated resonant annihilation into $\bar{f}f$}
\end{subfigure}
\begin{subfigure}{.25\textwidth}
  \centering\vspace{.3cm}
  \includegraphics[width=1.0\linewidth]{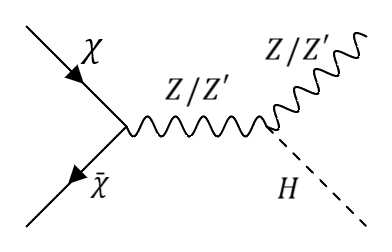}\vspace{.8cm}
  \caption{Higgstrahlung annihilation}
\end{subfigure}
\caption{Dark matter annihilation channels for the determination of the freeze-out relic density. }
\label{fig:dmchannels}
\end{figure}

\begin{figure}
\centering
  \includegraphics[width=.8\linewidth]{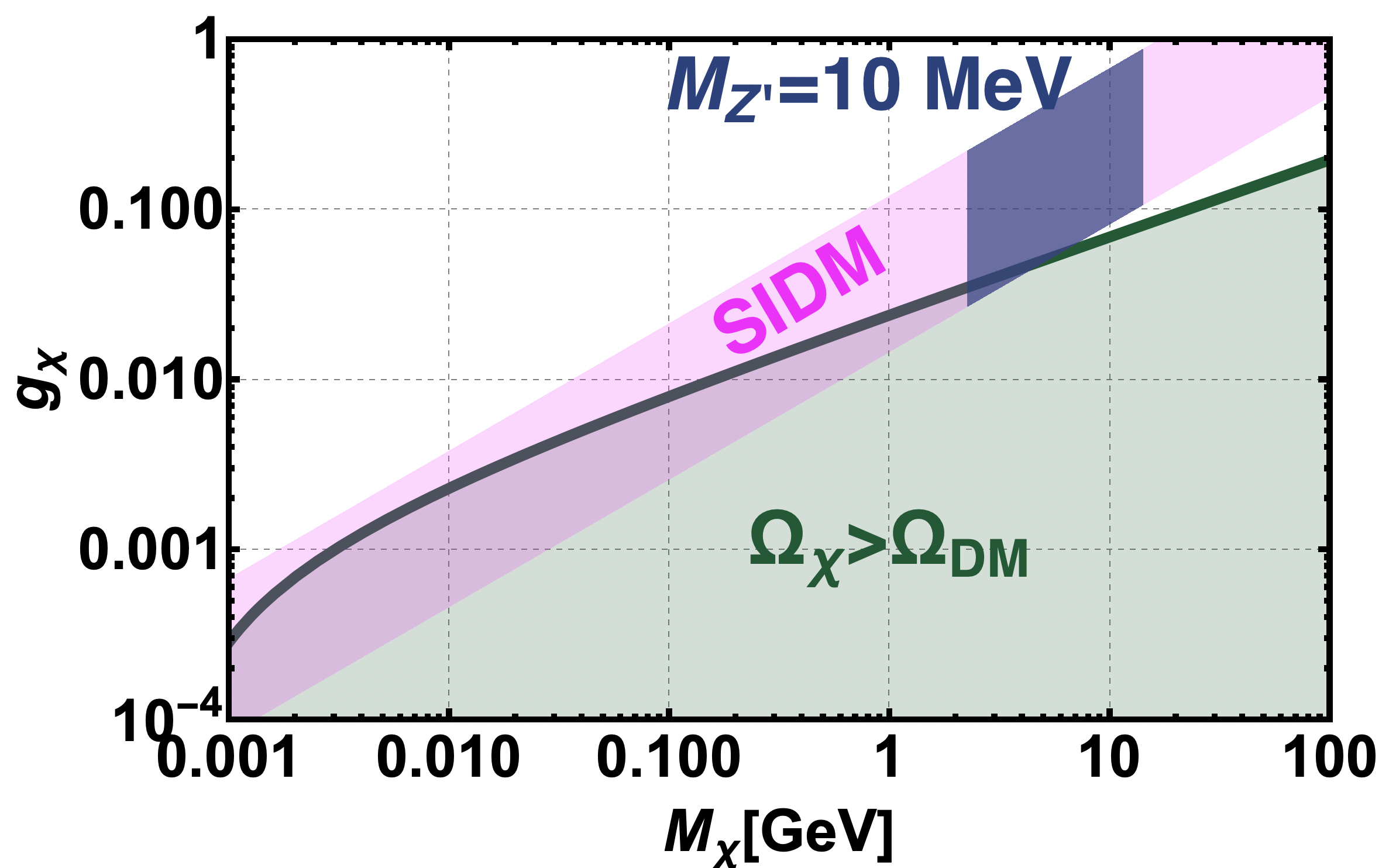}
\caption{Case 1: t-channel annihilation into a $Z'$ pair. Parameter space in the $M_\chi-g_\chi$ plane (where $ g_\chi = \gd\, \cos\theta_{\tt X}\approx \gd$ ) excluded by relic density overabundance. Area shaded in green results in relic density overabundance, while the green line corresponds to $\Omega_{DM}=0.1195$. The area indicated in Magenta with the label "SIDM" corresponds to the region where dark matter self-interactions are consistent with astrophysical observations. Within this area, the blue region corresponds to a mediator mass $M_{Z'}$ of 10 MeV, considering the uncertainty as estimated in the text. }
\label{fig:tchannelrelic}
\end{figure}

We consider the thermal freeze-out to determine the DM relic density, $\Omega_\chi$. We identify three scenarios which can result in a relic density of dark matter in accordance with cosmological measurements, $\Omega_\chi h^2 \leq 0.1198$:
\begin{enumerate}
     \item When $M_\chi>M_{Z'}$, the annihilation t-channel, $\bar{\chi} \chi\rightarrow Z' Z'$, is kinematically allowed. This leads to a relic density which only depends on the $Z'$ boson mass $M_{Z'}$, the dark gauge coupling $\gd$, and the dark matter mass $M_\chi$. Numerically we find that for each value of $M_{Z'}$ there is a minimum value of $\gd$ for which there is no dark matter overabundance. We show this behavior in Figure \ref{fig:tchannelrelic}. This scenario is well-studied and is known in the literature as Secluded WIMP Dark Matter 
      \cite{Pospelov:2007mp,Pospelov:2008jd,Batell:2009yf}.
     \item When $M_\chi>M_f$, $f$ being a SM fermion, the Z-Z' mixing allows the s-channel annihilation of dark matter into $f$, $\bar{\chi} \chi\rightarrow \bar{f} f$. In the resonant regions $M_\chi\sim M_{Z/Z'}/2$ the annihilation cross section can be enhanced enough to reach the required value to result in an allowed relic density, while keeping the value of $\theta_{\tt X}$ in the allowed region discussed in section \ref{sec:Zprimeconstraints} \cite{delaVega:2021wpx}. In this channel in addition to $M_{Z/Z'}$, $M_\chi$ and $\gd $, the Z-Z' mixing angle $\theta_{\tt X}$ is a crucial parameter. 
     \item When $2 M_\chi>M_{Z(Z')}+M_{S}$, where $S$ is one of the four neutral scalars in the model, the s-channel Higgsstrahlung channel ($\bar{\chi} \chi \rightarrow S Z(Z')$) is kinematically allowed. In this channel, the scalar masses and mixing angles become relevant to the relic density calculation. For dark matter masses above the $W$ boson mass, the channel $\bar{\chi} \chi\rightarrow W W$ is open and can contribute significantly to the dark matter relic density.
\end{enumerate}
We illustrate the tree-level Feynman diagrams of these processes in Figure \ref{fig:dmchannels}.
To calculate the relic density for this model, we have implemented the model in SARAH \cite{Staub:2013tta} and micrOmegas \cite{Belanger:2018ccd,Belanger:2020gnr}, scanning over a range of the free parameters. We study the first two cases, as they lead to an interesting interplay between the dark sector, the light gauge boson parameters and the neutrino sector.
\subsection{Dark matter direct detection}
\begin{figure}
\centering
  \includegraphics[width=.8\linewidth]{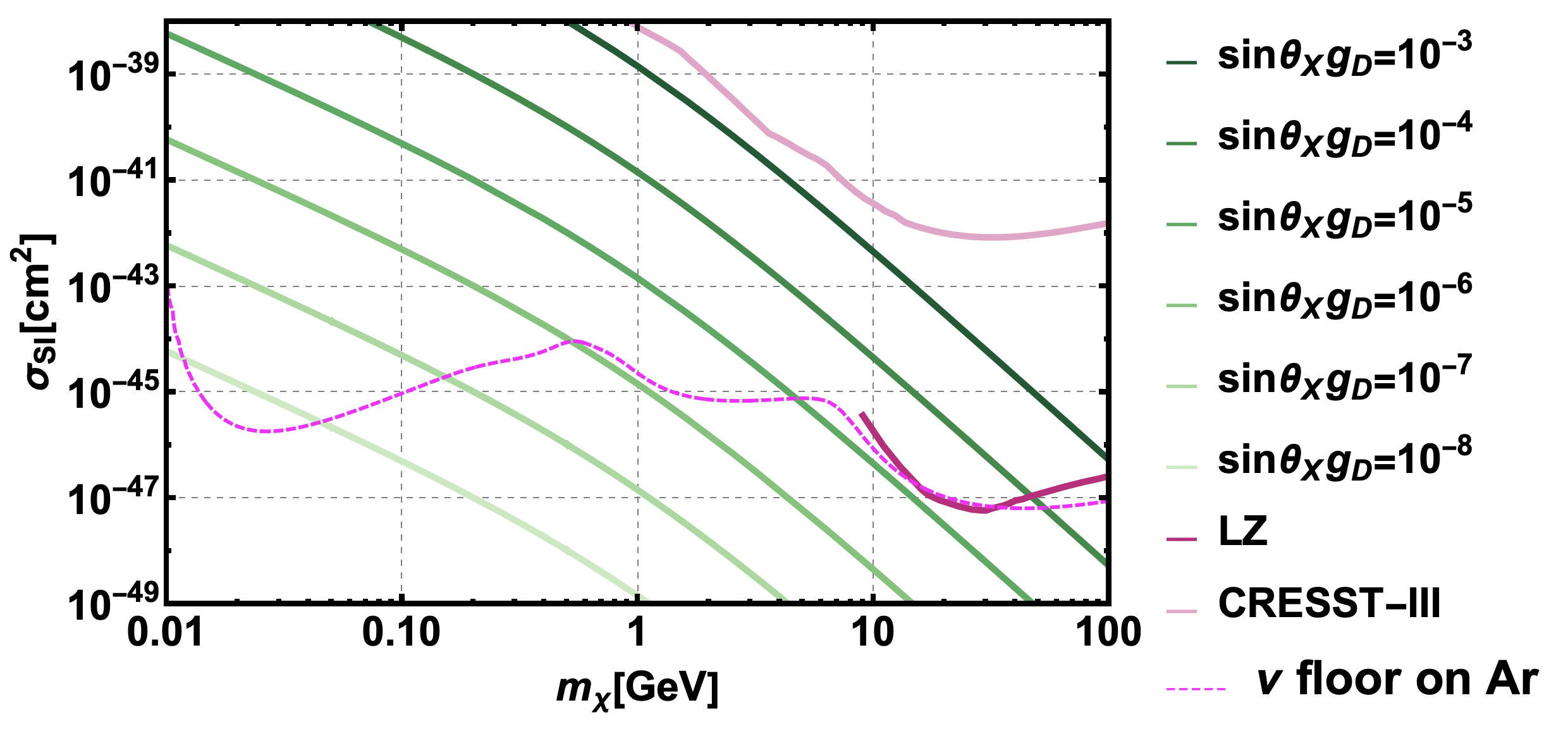}
\caption{Spin Independent direct detection cross section ($\sigma_{SI}$) as a function of dark matter mass, for the $Z'$ resonant case (s-channel). In the green lines the product $sin\theta_{\tt X} \gd $ is fixed to values between $10^{-8}$ and $10^{-3}$ as indicated in the caption. The purple lines show the experimental limit on $\sigma_{SI}$ set by the LZ \cite{LUX-ZEPLIN:2022qhg} and CRESST-III \cite{Abdelhameed_2019} experiments. The dashed magenta line shows the neutrino floor on Argon \cite{AristizabalSierra:2021kht}. Note that the LZ experiment target nucleus is Xenon, and it has not yet reached the Xenon neutrino floor.   }
\label{fig:resonantddcs}
\end{figure}
After excluding the parameter space where the relic density of $\chi$ does not correspond to that of dark matter, we look at the Spin-Independent cross section of dark matter with nucleons. The $Z-Z'$ mass mixing leads to tree-level dark matter-nucleon elastic scattering, a process searched for in direct detection experiments. The scattering is mediated by $Z$ and $Z'$ exchange. \\
In the case where relic density is determined by t-channel $\bar{\chi} \chi \rightarrow Z' Z'$ annihilations, there is no correlation between dark matter relic density and the direct detection cross section, as the relic annihilation cross section is independent of the $Z-Z'$ mixing angle at leading order. However, a bound on the Z-Z' mixing angle may be derived from the limits on this cross section. For the $M_{Z'}^2<<M_{Z}^2$ limit, the spin independent $\chi$-nucleus elastic scattering cross section is approximately \cite{Alves:2015mua}
\begin{equation}
\sigma_{SI}^{Z,A}=\frac{\mu^2_{\chi N}\sin^22\theta_{\tt X} g_X^2 Q_\chi^2    }{4\pi M_{Z'}^4} \left[ Z (2 g_{SM}^{Zu}+ g_{SM}^{Zd}) +(A-Z) ( g_{SM}^{Zu}+ 2g_{SM}^{Zd})  \right],
\end{equation}
where $\mu^2_{\chi N}$ is the reduced $\chi$-nucleus mass, $g_{SM}^{Zq}$ is the vector coupling of $q=u,d$ to the Z boson in the SM, and $Z$ and $A$ are the electric charge and atomic mass number of the nucleon respectively. Using the most stringent limits on dark matter direct detection we can list the following constraints 
\begin{itemize}
    \item $\frac{|g_X\sin2\theta_{\tt X}Q_\chi|}{2}\leq \sim 10^{-10}$ at $M_{Z'}=10$ GeV and $M_\chi=30$ GeV from LZ \cite{LUX-ZEPLIN:2022qhg}.
    \item $\frac{|g_X\sin2\theta_{\tt X}Q_\chi|}{2}\leq \sim 10^{-8}$ at $M_{Z'}=100$ MeV and $M_\chi=1$ GeV from CRESST-III \cite{Abdelhameed_2019}.
\end{itemize}
For the resonant Z' channel case we observe a clear correlation between the parameter product $\gd sin\theta_{\tt X}  $ and the SI cross section in Fig \ref{fig:resonantddcs}.
\subsection{Dark Matter Self-Interactions}
The dark gauge interaction provides a possible avenue for dark matter self-interactions, which can account for several contradictions between astronomical observations and the collisionless cold dark matter paradigm \cite{Tulin:2017ara}. Self-Interacting Dark Matter (SIDM) can explain the \textit{too big to fail}, \textit{core-cusp}, \textit{diversity} and \textit{missing satellites} problems.
\def\bP{\beta_\chi}
\def\mfe{M_\chi}
\def\mv{M_{Z'}}
Dark matter self-interactions in this model consist of $\chi \chi \rightarrow\chi \chi$ scatterings mediated by the $Z'$ boson. Defining
\begin{equation}
\beta_\chi = \sqrt{1 - \frac{4 M_\chi^2}s}\,, \quad g_\chi = \gd \cos\theta_{\tt X},
\end{equation}
we obtain the self-interaction cross section $\sigma_{\tt SIDM}$ \cite{Lamprea:2019qet}
\begin{align}
\frac{\sigma_{\tt SIDM}}\mfe&=
\frac{g_\chi ^4}{4 \pi s \mfe } \Biggl\{ 
\frac{(2s + 3\mv^2) s \bP^2+ 2(\mv^2+2\mfe^2)^2}{2\mv^2(\mv^2+s \bP^2)} \cr
&\hspace{1in}-\frac{  (s   \bP^2+ 2\mv^2) (3\mv^2+ 4\mfe^2) +2(\mv^2+2\mfe^2)^2- 4 \mfe^4  }{s\bP^2 \left(2 \mv^2 + s\bP^2\right)}\ln\left(1+\frac{s \bP^2}{\mv^2}\right)\Biggr\}\,;\cr
\end{align}
The requirements for SIDM are met in this model when this cross section is enhanced by the kinematic condition $ \mfe \gg \mv $, in the small relative velocity $ \bP$ regime.The SIDM cross section magnitude is determined by astrophysical data, namely dwarf and low surface brightness galaxies. A velocity-dependence of the cross section is obtained by the different typical velocities of dark matter in each environment. The central values of the cross sections and velocities are~\cite{Kaplinghat:2015aga} 
\begin{equation}
\left. \frac{\sigma_{\tt SIDM}}\mfe \right|_{\rm galaxy} \hspace{-.25in}=  1.9 \frac{\text{cm}^2}{\text{gr}} \,, ~ 
\left. \frac{\sigma_{\tt SIDM}}\mfe \right|_{\rm cluster}\hspace{-.25in} = 0.1 \frac{\text{cm}^2}{\text{gr}} \,;\quad
\left. \bP \right|_{\rm galaxy} =  3.3 \times 10^{-4} \,,~ 
\left. \bP \right|_{\rm cluster} = 5.4 \times 10^{-3}\,.
\end{equation}
Fitting to these values we find
\begin{equation}
\mv = \frac\mfe{566}\,, \qquad g_\chi = \left( \frac\mfe{75\, \text{GeV}} \right)^{3/4}\,.
\end{equation}
 We estimate the errors on the numerical coefficients of these expresions to be a factor of 2.5 (e.g., that the first coefficient ranges from $566/2.5$ to $2.5*566$). We find that these conditions can be met when dark matter relic density is obtained through annihilations to $Z'$ pairs. In Figure \ref{fig:tchannelrelic} we show the band where SIDM is viable. We note that there is an overlap between the SIDM band and the line where $\chi$ accounts for dark matter completely. In this scenario, the mass of the $Z'$ is of order $\sim 10$ MeV, which is where low energy experiments are most sensitive.
\subsection{Neutrino masses and $U(1)_D$ breaking scale}
\label{sec:neutrinomasspheno}
From the constraints on the dark sector parameters from the dark matter phenomenology, we can infer the following possibilities for the neutrino mass mechanism.\\
For all dark matter relic density channels, the inverse seesaw-like scenario 
\begin{equation}
        m_\nu\sim \frac{(m_D^2)^2 Y_{N'}}{M_1 Y_N},
    \end{equation}
and the type-I seesaw scenario
\begin{equation}
        m_l\sim \frac{v_1^2}{M_F},
    \end{equation}
are viable with heavy neutrinos of canonical seesaw scale.\\
Of special interest are the cases where the neutrino masses are linked to the $U(1)_D$ breaking scale, namely the limits considered in eqs. (\ref{seesaw1}) and (\ref{seesaw2}). In the limit considered in eq. (\ref{seesaw2}) 
 \begin{equation}
        m_\nu\sim \frac{(m_D^2)^2 Y_{N'}v_\phi^2}{M_1^2 M_F},
    \end{equation}
a light neutrino masses of the order $\mathcal{O}(\text{eV})$ can be achieved with $\mathcal{O}(\text{TeV})$ heavy neutrinos, as in the canonical inverse seesaw. In this way the smallness of neutrino masses is linked to the smallness of the $Z^\prime$ boson mass.\\

Lets consider now the limit in eq. (\ref{seesaw1}). For the t-channel annihilation case, the correct relic density is determined for values of $\gd$ larger than $3\times 10^{-4}$ for a dark matter mass of $\sim1$ MeV, or $\gd$ larger than $2\times 10^{-1}$ for a dark matter mass of $\sim100$ GeV (see Figure \ref{fig:tchannelrelic}). In the light $Z'$ paradigm, with small $Z-Z'$ mixing we have
\begin{equation}
    M_{Z'}>\gd v_\phi,
\end{equation}
and the t-channel dominated annihilation has the kinematic condition
\begin{equation}
    M_\chi>M_{Z'}.
\end{equation}
These two equations rule out the inverse seesaw-like limit of neutrino masses
\begin{equation}
        m_\nu\sim \frac{(m_D^1)^2 M_1}{v_\phi^2 Y_N Y_{N'}},
    \end{equation}
as the low scale of $v_\phi$ needed to obtain a light $M_{Z'}$ would result in either $\sim 1 \text{GeV}$ scale sterile neutrinos with large mixings with the active neutrinos, or neutrino Yukawa couplings much smaller than the electron Yukawa coupling of the SM, calling into question the necessity of the seesaw scheme.
\section{Conclusions}
Abelian gauge extensions of the SM with light gauge mediators have become popular recently because they lead to interesting implications in low energies experiments and observables. In this work we have studied a scenario where the coupling of such a light gauge boson with the SM is generated through a mass mixing with the $Z$ boson. We extended the model by including a DM fermion charged under the new $U(1)_D$ symmetry which is automatically stable without the inclusion of extra symmetries. In this way the DM phenomenology further constrains the new gauge sector. The DM self-interactions are also discussed and we present the parameter space where the cusp-core problem is resolved.
Focusing on the mass region for the $Z'$ gauge boson where low energy parity violation experiments are more sensitive, $M_Z\sim 10 ~\text{MeV}$ we have found the parameter space for the dark gauge coupling, $g_D$ and mass mixing parameter to reproduce the DM relic abundance and search for the region where the DM self-interaction explain the approximately constant dark matter density in the inner parts of galaxies. We have found an overlap to explain correctly the DM relic abundance and the DM self-interaction in the DM mass region $\sim 2-7 ~\text{GeV}$ with $g_D$ above $\sim 0.01$. We also identify a scenario where there is no significant DM self-interactions, but DM can be seen in direct detection experiments.
The neutrino masses are generated through the seesaw mechanism. In some cases, the $U(1)_D$ breaking scale plays a crucial role in the neutrino mass generation.  In these scenarios it is possible to generate neutrino masses through a low-energy seesaw.
\appendix
\section{Tree-level stability of the potential}
\label{sec:AppendixScalarStability}
Let
\begin{equation}
    x_1 = \sqrt{2}\,|H_1|^2\,,\quad x_2 = \sqrt{2}\,|H_2|^2\,, \quad x_3 = \sqrt{2}\,|\phi|^2,
\end{equation}
then the quartic part of the potential takes the form
\begin{equation}
    V_4 = \frac12 \sum_{i=1}^3 \lambda_i x_i + \eta_3 x_1 x_2 + \eta_1 x_2 x_3 + \eta_2 x_3 x_1,
\end{equation}
where $\eta_1 = \lambda_5/2,\, \eta_2 = \lambda_6/2,\,\eta_3 = (\lambda_4+\zeta^2\lambda_7)/2$ with $ \zeta = |H^\dagger_1 H_2|/(|H_1|\, |H_2|)$ (note that $ 0 \le\zeta\le1$). The tree level stability conditions are then: $ \lambda_i>0$ and $ \eta_i > - \sqrt{\lambda_j \lambda_k}$ with $\{i,\,j,\,k\}$ a cyclic permutations of $\{1,\,2,\,3\}$; in addition,
\begin{itemize}
    \item If $ \eta_i>0\,,~\,\eta_j,\,\eta_k<0$, then $\lambda_i \eta_i > \eta_j \eta_k  - \sqrt{(\lambda_i \lambda_j-\eta_k^2)(\lambda_i \lambda_k-\eta_j^2)} $.
    \item If $ \eta_{i,j,k}<0$ then $\lambda_1 \lambda_2\lambda_3 + 2 \eta_1\eta_2\eta_3 > \lambda_1 \eta_1^2 + \lambda_2\eta_2^2 + \lambda_3\eta_3^2 $.
\end{itemize}

\section*{Acknowledgements}
The authors thank Luis J. Flores for kindly providing the neutrino floor data and R. Ferro-Hernandez for useful discussions. LMGDLV thanks CONACYT for the funding of his PhD studies. This work has been supported by the University of California Institute for Mexico and the United States (UC MEXUS) (CN 18-128) and the Consejo Nacional de Ciencia y Tecnología (CONACYT) (CN 18-128), by the German-Mexican research collaboration grant SP 778/4-1 (DFG) and 278017 (CONACYT), the CONACYT CB-2017-2018/A1-S-13051, and  DGAPA-PAPIIT IN107621.
\bibliography{apssamp}

\end{document}